\documentclass[12pt,preprint]{aastex}
\usepackage {graphicx}
\usepackage{aastexug}

\shorttitle{Correlation between 3:2 QPO pairs and Jets in Black Hole
X-ray Binaries} \shortauthors{D. X. Wang et al.}

\begin{document}

\title{Correlation between 3:2 QPO pairs and Jets in Black Hole X-ray Binaries}

\author{Ding-Xiong Wang\altaffilmark{1},  Yong-Chun Ye and Chang-Yin Huang}
\affil{Department of Physics, Huazhong University of Science and
Technology,Wuhan, 430074, P. R. China}

\altaffiltext{1}{Send offprint requests to: D. X. Wang
(dxwang@hust.edu.cn)}

\begin{abstract}

We argue, following our earlier works (the `CEBZMC model'), that the
phenomenon of twin peak high frequency quasi-periodic oscillations
(QPOs) observed in black hole X-ray binaries is caused by magnetic
coupling (MC) between accretion disk and black hole (BH). Due to MC,
two bright spots occur at two separate radial locations $r_{in}$ and
$r_{out}$ at the disk surface, energized by a kind of the
Blandford-Znajek mechanism (BZ). We assume, following the
Kluzniak-Abramowicz QPO resonance model, that Keplerian frequencies
at these two locations are in the 3:2 ratio. With this assumption,
we estimate the BH spins in several sources, including GRO J1655-40,
GRS 1915+105, XTE J1550-564, H1743-322 and Sgr A*. We give an
interpretation of the `jet line' in the hardness-intensity plane
discussing the parameter space consisting of the BH spin and the
power-law index for the variation of the large-scale magnetic field
in the disk. Furthermore, we propose a new scenario for the spectral
state transitions in BH X-ray binaries based on fluctuation in
densities of accreting plasma from a companion star.

\end{abstract}


\section{INTRODUCTION }

Data collected by the NASA satellite Rossi X-Ray Timing Explorer
(RXTE; Bradt, Rothschild \& Swank 1993) added a new impetus to
studies of QPOs, observed in X-ray binaries and other sources. The
QPO observations are described in several recent reviews, e.g. by
Remillard (2005) or Remillard \& McClintock 2006 (hereafter RM06).
The QPO observations present several puzzles, including why is the
occurrence of high frequency QPOs correlated with the occurrence of
relativistic jets. The jets in microquasars were first observed by
Mirabel \& Rodrigues (1998, 1999). For references to more recent
works see McClintock \& Remillard (2006, hereafter MR06) and Kalemci
et al. (2006). It is widely agreed (Mirabel \& Rodrigues 1999;
Blandford 2002) that in both active galactic nuclei (AGNs) and
microquasars, relativistic jets must be powered by a process similar
to the celebrated Blandford-Znajek (1977, hereafter BZ) mechanism. A
particular variant of BZ, in the form of magnetic couplings (MC)
between a rotating BH and its surrounding disk, has been recently
investigated by several authors, including Blandford (1999), Li
(2000, 2002), and Wang et al. 2002, hereafter W02). Our `CEBZMC'
model  (Wang et al. 2003a, 2004, hereafter W04) also belongs to the
BZ plus MC class.

In this paper we consider a particular realization of the CEBZMC
model, in which MC between BH and accretion disk energizes `hot
spots' on the accretion disk surface. As discussed by Wang et al.
2005 (hereafter W05), a pair of hot spots produce a pair of QPOs,
with frequencies $\nu_{in}$ and $\nu_{out}$ that correspond to
Keplerian frequencies at the locations of the two spots, $r_{in}$
and $r_{out}$. We \textit{assume} that the two locations agree with
the Kluzniak-Abramowicz resonance condition, $\nu_{in} /
\nu_{out}=3/2$. Abramowicz and Kluzniak (2001, hereafter AK01)
realized that frequencies of the QPO pairs in BH sources are in the
exact 3:2 ratio. They also recognized and stressed the fundamental
importance of this fact\footnote{Abramowicz et al. (2003) also
argued that~there is a statistical evidence for the same 3:2 ratio
for QPOs in neutron star sources. This was later confirmed by
Belloni et al. (2005). More recently, Abramowicz et al. (2005) found
an additional, independent, and direct proof for the 3:2 ratio in
the QPO neutron star data. They proved that although in neutron star
sources the \textit{observed} frequencies vary in a wide range, the
variations uniquely point to the 3:2 ratio of the
\textit{eigenfrequencies}.}. They first noticed the 3:2 ratio in the
QPO pair with frequencies 450 Hz and 300 Hz in GRO J1655-40,
observed by Strohmayer, (2001a, b). This important discovery was
strengthened by numerous authors, who found the 3:2 ratio in
different BH sources: in GRS 1915+105 MR06, in XTE J1550-564 Miller
(2001) and Remillard et al (2002), and in H1743-322 Homan et al.
(2005) and Remillard et al. (2006). There is less certain evidence
that the same 3:2 ratio occurs for QPOs observed in low-mass active
galactic nuclei - in e.g. Sgr A* (T$\ddot{o}$r$\ddot{o}$k 2005a, b;
Ashenbach 2005) and in a few nearby Seyferts (Lachowicz et al.,
2006). The 3:2 ratio of frequencies is the basic feature of the
Kluzniak-Abramowicz QPO resonance model (see a collection of reviews
in Abramowicz 2005 for references). In other QPOs models, the 3:2
ratio of observed frequencies is either incidental or impossible as
for example, in the Lamb and Miller (2004) model that considered the
QPO pairs to be a beat frequency between neutron star spin and disk
rotation, or in the Wagoner et al., (2001) model, in which they are
fundamental g-mode and c-mode in thin disk oscillations, or in the
Stella and Vietri (1999) model where they emerge as a combination of
Keplerian and radial epicyclic frequencies. In our CEBZMC model, in
its present state of development, the 3:2 ratio does not directly
follow from the model basic assumption (MC, BZ). However, as we will
see later, the 3:2 ratio occurs in an interesting region of the
parameter space of the CEBZMC model. We therefore \textit{assumed}
that the ratio is equal 3:2, and examined consequences of this
assumption. This phenomenological approach has at least three
virtues. Firstly, it allows us to estimate the spin of BHs in the BH
sources that display the QPOs with the 3:2 ratio, using a method
first applied by AK01 and more recently T$\ddot{o}$r$\ddot{o}$k et
al. (2005). Secondly, it is directly related to the correlation
between the occurrence of the QPO 3:2 pairs and the jets. Thirdly,
we may offer an interpretation of the 'jet line'  in the
hardness-intensity diagram (HID), by which the hard state and the
soft state of the BH X-ray binaries (BHXBs) are separated (Fender et
al. 2004, hereafter FBG04; Belloni 2006, hereafter B06, Remillard
2005).

This paper is organized as follows. In $\S$ 2 we give a brief
description of our model and explain the correlation between the 3:2
QPO pairs and the jets in BHXBs. In $\S$ 3 we compare the BH spins
measured by different methods. It turns out that the BH spin of GRO
J1655-40 estimated by CEBZMC is consistent with those estimated by
X-ray continuum fittings. By fitting the 3:2 QPO pairs, we estimate
the spin of the galactic massive BH, Sgr A*, for a given range of
the BH mass, and also estimate the spin and mass of the BH candidate
H1743-322. In addition, the spin of the BH X-ray binary XTE
J1859+226 is estimated by fitting its single--component high
frequency QPO. In $\S$ 4 we propose a new scenario for the state
transitions in BHXBs based on the variation of the power-law index
$n$, which arises from the fluctuation of the number densities of
the accreting plasma from a companion star. Finally, in $\S$ 5, we
summarize the main results, and discuss the issues related to this
model. Throughout this paper the geometric units $G = c = 1$ are
used.

\section{CORRELATION BETWEEN 3:2 QPO PAIRS AND JETS }

In order to discuss the correlation of the 3:2 QPO pairs with the
jets in the BHXBs we give a brief review of our previous works. In
W05 we approached the 3:2 QPO pairs by virtue of the MC of a Kerr BH
with its surrounding disk as shown in Figure 1, in which the
large-scale magnetic field at the BH horizon consists of the open
and closed field lines with an angular boundary at $\theta _S $. The
open field lines transfer the energy and angular momentum from the
BH to the remote astrophysical loads in the BZ process, while the
closed field lines transfer those between the BH and the surrounding
accretion disk in the MC process. The angular boundary $\theta _S $
is determined by a criterion of the screw instability of the
magnetic field given in W04.

The upper and lower frequencies of the 3:2 QPO pairs correspond
respectively to the inner and outer hotspots rotating with the
Keplerian angular velocities of the disk, which are produced by the
MC with the non-axisymmetric magnetic field at the BH horizon (Wang
et al. 2003b). As argued in W05, the positions of the inner and
outer hotspots are determined by the maximum radiation flux from the
disk and the screw instability of the non-axisymmetric magnetic
field, respectively.

It turns out that the 3:2 QPO pairs fitted in our model depend
mainly on two parameters, i.e., the BH spin $a_ * $ and the
power-law index $n$. The parameter $a_ * \equiv J \mathord{\left/
{\vphantom {J {M^2}}} \right. \kern-\nulldelimiterspace} {M^2}$ is
defined in terms of the BH mass $M$ and angular momentum $J$, and
the parameter $n$ is defined in terms of the variation of the
poloidal magnetic field $B_D^P $ with the disk radius, $B_D^P
\propto r^{ - n}$.

It has been argued in W04 that the state of CEBZMC always
accompanies the screw instability, provided that the BH spin $a_ * $
and the power-law index $n$ are greater than some critical values.
Based on the criterion of the screw instability derived in W04 we
have a contour of the angular boundary $\theta _S \left( {a_ * ,n}
\right) = 0$ in the $a_ * - n$ parameter space as shown in Figure 2.

Inspecting Figure 2, we have the following results:

(\ref{eq1}) The shaded region indicated ``BZMC with Jet'' represents
the value ranges of the parameters $a_ * $ and $n$ for CEBZMC, in
which the jet driven by the BZ process exists.

(\ref{eq2}) The inner hotspot arises from energy transferred from a
fast-rotating BH into the disk by non-symmetric MC.

(\ref{eq3}) The outer hotspot is produced by the screw instability,
which always accompanies the state of CEBZMC.

Recently, the 3:2 QPO pair has been observed in near infrared flares
of the massive BH Sgr A* in the Galactic Center (1.445, 0.886mHz;
Aschenbach 2004a, 2004b; T$\ddot{o}$r$\ddot{o}$k 2005). In addition,
Bower et al. (2004) state that their radio measurements of Sgr A*
are consistent with jet models. Based on the model of CEBZMC given
in W04 and W05 we have the fitting results of the 3:2 QPO pairs for
the three BHXBs and Sgr A* as shown in Table 1. It should be noticed
that the BH spins given in Table 1 are a little less than those
given in W05 (see Table 1), because some errors in calculations have
been corrected and the ranges of the BH masses have been updated.

As argued in W05, the upper and lower frequencies of the 3:2 QPO
pairs are equal to the Keplerian frequencies of the inner and outer
hotspots, respectively, being expressed by

\begin{equation}
\label{eq1} \nu _i = \nu _0 (\xi _i^{3 \mathord{\left/ {\vphantom {3
2}} \right. \kern-\nulldelimiterspace} 2} \chi _{ms}^3 + a_ * )^{ -
1},
\end{equation}

\noindent where $\nu _0 \equiv \left( {m_{BH} } \right)^{ - 1}\times
3.23\times 10^4Hz$ with $m_{BH} \equiv M \mathord{\left/ {\vphantom
{M {M_ \odot }}} \right. \kern-\nulldelimiterspace} {M_ \odot }$.
The parameter $\xi _i \equiv {r_i } \mathord{\left/ {\vphantom {{r_i
} {r_{ms} }}} \right. \kern-\nulldelimiterspace} {r_{ms} }$ is the
disk radius expressed in terms of $r_{ms} $, the radius of the
innermost stable circular orbit (ISCO). The frequency $\nu _i $
represents $\nu _{upper} $ and $\nu _{lower} $ for $\xi _i $ equal
to $\xi _{upper} $ and $\xi _{lower} $, respectively. As argued in
W05 both $\xi _{upper} $ and $\xi _{lower} $ depend on the parameter
$a_
* $ and $n$, i.e.,

\begin{equation}
\label{eq2} \left\{ {\begin{array}{l}
 \xi _{upper} = \xi _{upper} \left( {a_ * ,n} \right), \\
 \xi _{lower} = \xi _{lower} \left( {a_ * ,n} \right). \\
 \end{array}} \right.
\end{equation}

It is obvious that the 3:2 QPO pair can be completely determined by
combining equation (\ref{eq2}) with equation (\ref{eq1}), provided
that the BH mass $m_{BH} $, $\nu _{upper} $ and $\nu _{lower} $ are
given. This implies that the 3:2 QPO pair with the given BH mass
corresponds to one `representative point' in the $a_ * - n$
parameter space. Thus we have a characteristic line of the 3:2 QPO
pair for a continuous distribution of $m_{BH} $ within its upper and
lower limits as shown by the thick solid line in Figure 3.

On the other hand, $\nu _{upper} = const$ corresponds to a contour
with the given BH mass in the $a_ * - n$ parameter space based on
equations (\ref{eq1}) and (\ref{eq2}), and we have two contours of
$\nu _{upper} = const$ corresponding to the lower and upper BH
masses in the parameter space as shown in Figure 3.

It is found from Figure 3 that the characteristic lines of the 3:2
QPO pairs for the four BH systems are all located in the shaded
region indicated ``BZMC with Jet''. These results provide a natural
explanation for the correlation between the 3:2 QPO pairs and the
jets driven by the BZ process, being consistent with the fact that
jets are found in the above BH systems (Mirabel {\&} Rodrigues 1998,
1999; Aschenbach 2004b; Bower et al. 2004; T$\ddot{o}$r$\ddot{o}$k
2005; MR06).

\section{ESTIMATING BH SPINS BY VIRTUE OF HIGH FREQUENCY QPOS }

A Kerr BH is described completely by its mass $ M$ and spin $a_ * $.
The masses of twenty BHs in the Galaxy have already been measured or
constrained, and the next goal is to measure spin. As pointed out by
RM06, there are four avenues for measuring BH spin, which include
(\ref{eq1}) X-ray polarimetry, (\ref{eq2}) X-ray continuum fitting,
(\ref{eq3}) the Fe K line profile and (\ref{eq4}) high frequency
QPOs. Among these approaches high frequency QPOs are likely to offer
the most reliable measurement of spin once the correct model is
known. Unfortunately, there are significant differences in the BH
spins measured by different models, and a reasonable model for
measuring BH spins has not been accepted by astrophysical community.
In this paper we compare the values of the BH spins of the three
BHXBs and Sgr A*, which are measured by X-ray continuum and 3:2 QPO
pairs as listed in Table 2.

The method of X-ray continuum fitting is used to measure the BH
spins of the binaries based on a fully relativistic model of a thin
accretion disk around a Kerr BH. In order to estimate the BH spin by
fitting the broadband X-ray spectrum, one must know the BH mass, the
inclination$ i$ of the accretion disk, and the distance to the
binary.

The approach to the BH spin based on the 3:2 QPO pair consists of
two basic methods. One method is based on the epicyclic resonance
model (ERM), in which the resonance between orbital and epicyclic
motions of accreting matter is invoked (Abramowicz {\&} Kluznick
2004 and references therein), and another method is based on CEBZMC,
in which the inner and outer hotspots are produced by a
non-axisymmetric MC and the screw instability of the magnetic field,
respectively. The BH spin measured by ERM and CEBZMC depends on the
BH masses.

From Table 2 we find that the spin of GRO J1655-40 measured by
CEBZMC is in a good agreement with those measured by X-ray continuum
fittings given in G01, S06 and M06. However, an intersection of the
BH spins has not been found for XTE J1550-564 and GRS 1915+105 based
on the above two methods.

It is found that the spin of Sgr A* measured by CEBZMC is generally
not overlapped with those measured by ERM except those given in
Br05. Up to date, the BH spin of Sgr A* has been estimated only by
using ERM and CEBZMC, depending sensitively on the BH mass. For
example, based on CEBZMC the spin is estimated as 0.811--0.951 and
0.800--0.841 for $m_{BH} = \left( {2.6 - 4.4} \right)\times 10^6$
and $\left( {2.53 - 2.84} \right)\times 10^6$, respectively. Based
on ERM the spin is constrained to be 0.9865--0.9965 and 0.99616 for
$m_{BH} = \left( {2.53 - 2.84} \right)\times 10^6$ and $3.3\times
10^6$ in A04a and A06, respectively.

A common feature in the above measurements lies in the fact that the
BH spins are constrained more tightly for the narrower ranges of the
BH masses. Although the BH mass of H1743-322 has not been
constrained, it is identified as a BH candidate by the X-ray light
curve and variability characteristics during its 2003 outburst, and
its behavior resembles the BHXBs XTE J1550-564 and GRO J1655-40 in
many ways (Remillard et al. 2002, 2006). It is interesting to note
that both the 3:2 QPO pair and the jet have been observed in
H1743-322 also (Homan et al. 2005; Remillard et al. 2006; Kalemci et
al. 2006). Thus we can constrain the BH mass and spin also by the
3:2 QPO pair (240, 160Hz) based on the model of CEBZMC.

As shown in Figure 3, a characteristic line in the $a_ * - n$
parameter space represents the 3:2 QPO pair, which is located
between two contours of $\nu _{upper} = const$ for the lower and
upper BH masses. In the case of H1743-322 the BH mass and spin can
be also constrained by the characteristic line above the contour
$\theta _S \left( {a_ * ,n} \right) = 0$ in the $a_ * - n$ parameter
space as shown in Figure 4.

Inspecting Figure 4, we find that the characteristic line is located
in the shaded region indicated by ``BZMC with Jet'', and it spans a
very wide range of the BH spin. Required by the 3:2 QPO pair (240,
160Hz) and the upper limit ($a_ * \le 0.998)$ to the BH spin given
by Thorne (1974) we have the leftmost and rightmost points of the
characteristic line as follows:

\begin{equation}
\label{eq3} \left( {a_ * ,n,\xi _{\max } } \right) = \left\{
{\begin{array}{l}
 (0.371,\mbox{ }5.670,\mbox{ }2.271), \\
 (0.998,\mbox{ }4.105,\mbox{ }1.187), \\
 \end{array}} \right.
\end{equation}

\noindent where the upper and lower lines correspond to the leftmost
and rightmost points of the characteristic line in Figure 4,
respectively. Combining $\nu _{upper} = 240Hz$ with equations
(\ref{eq1})--(\ref{eq3}) for the leftmost and rightmost points of
the characteristic line, we can estimate the value range of the BH
mass: $3.76 < m_{BH} < 48.23$. Although the BH mass and spin of
H1743-322 are only constrained loosely by the 3:2 QPO pair, they can
be further constrained by other observations. For example, the BH
mass and spin can be limited to a smaller range by fitting the
observed jet power in terms of the BZ power based on the model of
CEBZMC.

It is well known that single--component high frequency QPOs have
been observed in some confirmed and candidate BHXBs (MR06), such as
XTE J1859+226 (190 Hz), 4U1630--47 (184 Hz) and XTE J1650--500 (250
Hz). According to the model of CEBZMC the single--component high
frequency QPO can be fitted by ONE rotating hotspot arising from the
maximum radiation flux due to the non-axisymmetric MC.

Since neither jets nor 3:2 QPO pairs are observed in XTE J1859+226,
its state should be confined in the shaded region below the contour
of $\theta _S \left( {a_ * ,n} \right) = 0$ as shown in Figure 5. As
argued by Li (2002) the minimum spin for transferring energy from
the BH to the disk in the MC process is $a_ * = 0.3594$, and it is
regarded as the left boundary of the shaded region in Figure 5. Thus
the BH spin of XTE J1859+226 can be estimated as $0.3594 < a_ * <
0.5890$ by combining $\nu _{QPO} = 190Hz$ with the BH mass, $7.6 <
m_{BH} < 12.0$, which is taken from RM06.

\section{A SCENARIO FOR STATE TRANSITIONS IN BHXBS }

The prominent feature of the model of CEBZMC lies in the correlation
of the high frequency QPO pairs with the jets from the BHXBs. As is
well known, state transitions in BHXBs involve a number of
unresolved issues in astrophysics, displaying complex variations not
only in the luminosities and energy spectra, but also in
presence/absence of jets and QPOs. How to analyze and classify
states in BHXBs from observations in multi-wavelength band is of
foremost importance.

Recently, FBG04 proposed a unified semi-quantitative model for the
disk-jet coupling in BHXBs, in which the states of BHXBs are
described in an X-ray hardness-intensity diagram (HID), and the
states with jet and those with no jet are divided by a 'jet line' in
HID. Later, B06 classified the states of BHXBs into four types:
(\ref{eq1}) Low/Hard State (LS), (\ref{eq2}) Hard Intermediate State
(HIMS), (\ref{eq3}) Soft Intermediate State (SIMS) and (\ref{eq4})
High/Soft State (HS), which display different luminosity and
hardness associated with different behavior of QPOs and radio
loudness. It is pointed out in B06 that these states might be
reduced to only two basic states, i.e., a hard state and a soft one.
The states LS and HIMS are included in the hard state, and the
states SIMS and HS in the soft state. The jets can be observed in
hard states, but can not in soft states.

Very recently, MR06 used four parameters to define X-ray states
based on the very extensive RXTE data archive for BHXBs, in which
three states are included: (\ref{eq1}) thermal state (high/soft
state), (\ref{eq2}) hard state (low/hard state) and (\ref{eq3})
steep power law (SPL) state. In the thermal state, the flux is
dominated by the heat radiation from the inner accretion disk, and
QPOs are absent or very weak. The hard state is characterized by a
hard power-law component at 2--20 keV, being associated with the
presence of a quasi-steady radio jet. The SPL state is a strong
power-law component with $\Gamma $ $\sim $ 2.5, which is associated
with high-frequency QPOs. In MR06 luminosity is abandoned as a
criterion for defining the X-ray states.

However, a consistent interpretation for the state transitions in
BHXBs remains controversial, and this becomes a great challenge to
the present theoretical models. Some authors (e.g. Belloni et al.
1997a,b; 2000) interpreted the transition between State C and States
A/B as being caused by the disappearance and reappearance of the
inner accretion disk due to a disk instability mechanism. Livio et
al. (2003) pointed out that the inner disk remains present rather
than absent in the state transitions, and it switches between two
states in two different ways of converting accretion energy. In one
state, the accretion energy is dissipated locally to produce the
observed disk luminosity. In another state the energy liberated in
the accretion is converted efficiently into magnetic energy in the
form of a magnetically dominated outflow or jet. However, a detailed
argument for producing jets in BHXBs has not been given by these
authors.

Motivated by the above discussion we suggest a new scenario for the
state transition in BHXBs based on the model of CEBZMC. Inspecting
Figures 2--4, we find that the two basic states suggested by B06 can
be naturally divided by the contour of $\theta _S \left( {a_ * ,n}
\right) = 0$ in $a_ * - n$ parameter space: a hard state with jet is
represented by a point in the shaded region above the contour of
$\theta _S \left( {a_ * ,n} \right) = 0$, while a soft state without
jet by a point in the region below this contour. The state
transition in BHXBs can be interpreted in terms of the variation of
the power-law index$ n$. As shown in Figure 6, a hard state will
transit to a soft state with the decreasing $n$, while a soft state
will change to a hard state with the increasing $n$. The contour of
$\theta _S \left( {a_ * ,n} \right) = 0$ corresponds exactly to the
`jet line' in HID.

One of the main problems of this scenario is in knowing what
mechanism gives rise to the variation of the power-law index $n$.
This issue might be related to the fluctuation in the number density
of the accreting plasma from the companion star, and a rough
explanation is given as follows.

In our model the power-law index $n$ is used to describe the
variation of the poloidal magnetic field with the disk radius, i.e.,

\begin{equation}
\label{eq4} B_D^p = \left( {B_D^p } \right)_{ms} \left( {r
\mathord{\left/ {\vphantom {r {r_{ms} }}} \right.
\kern-\nulldelimiterspace} {r_{ms} }} \right)^{ - n} = \left( {B_D^p
} \right)_{ms} \xi ^{ - n},
\end{equation}

\noindent where $\left( {B_D^p } \right)_{ms} $ is the poloidal
magnetic field at ISCO. Based on Ampere's law we have the toroidal
current density $j_\varphi $ at the disk as follows,

\begin{equation}
\label{eq5} j_\varphi = \frac{1}{4\pi }\frac{dB_D^p }{dr} =
\frac{1}{4\pi r_{ms} }\frac{dB_D^p }{d\xi } = - \frac{n\left( {B_D^p
} \right)_{ms} }{4\pi M\chi _{ms}^2 }\xi ^{ - \left( {n + 1}
\right)}.
\end{equation}

From equations (\ref{eq4}) and (\ref{eq5}) we find that the profile
of the magnetic field at the disk is related directly to the
toroidal current at the same place, and the power-law index of the
latter becomes $n + 1$. Thus the variation of the magnetic field can
be explained by the variation of the toroidal current, and the
latter might be produced due to the fluctuation of the accreting
plasma coming from the companion star.

As a simple analysis, we assume that the accreting plasma consists
of electrons and protons, of which the number densities are $n_e $
and $n_p $, respectively. Generally, the two number densities are
not equal exactly, and they are related by $n_p = n_e + n_\delta $.
Thus a toroidal current density could be generated due to the
charged particles' Keplerian rotation and it reads

\begin{equation}
\label{eq6} j_\varphi = en_\delta \upsilon _\varphi = {en_\delta \xi
\chi _{ms}^2 } \mathord{\left/ {\vphantom {{en_\delta \xi \chi
_{ms}^2 } {\left( {\chi _{ms}^3 + a_ * } \right)}}} \right.
\kern-\nulldelimiterspace} {\left( {\chi _{ms}^3 + a_ * } \right)}.
\end{equation}

\noindent where $e = 4.8\times 10^{ - 10}e.s.u.$ is the electron
charge. Incorporating equations (\ref{eq5}) and (\ref{eq6}), we have

\begin{equation}
\label{eq7} n_\delta = - \frac{n\left( {B_D^p } \right)_{ms} }{4\pi
e\xi M}\left( {\frac{\chi _{ms}^3 + a_ * }{\chi _{ms}^4 }}
\right)\xi ^{ - \left( {n + 1} \right)}.
\end{equation}

As argued in W05, $B_4 \approx 10^5$ is the strength of the magnetic
field required by the hotspots for emitting X-ray. Taking $\left(
{B_D^p } \right)_{ms} = B_4 \times 10^5gauss$, $M = m_{BH} M_ \odot
$ and $\xi = 1$, we have

\begin{equation}
\label{eq8} \mid n_{\delta } \mid = 4.5\times 10^8\times \left( {B_4
m_{BH}^{ - 1} } \right)\left( {\chi _{ms}^{ - 1} + a_ * \chi _{ms}^{
- 4} } \right)cm^{ - 3}.
\end{equation}

By taking the disk mass as $M_{disk} = \alpha _m m_{BH} M_ \odot $,
the average disk height as $H = \beta r$ and the outer boundary
radius $r_{out} = 1000r_{ms} $, the average number density of
protons can be estimated as

\begin{equation}
\label{eq9} \bar {n}_p = \frac{M_{disk} }{m_p \int_{r_{ms}
}^{r_{out} } 2 \pi rHdr} = \left( {\alpha _m \beta ^{ - 1}m_{BH}^{ -
2} \chi _{ms}^{ - 6} } \right)\times 1.77\times 10^{32}cm^{ - 3},
\end{equation}

\noindent where $m_p = 1.67\times 10^{ - 24}g$ is a proton's mass.
Incorporating equations (\ref{eq8}) and (\ref{eq9}) with the given
values of the concerned parameter, such as $\alpha _m \approx 10^{ -
3}$, $\beta = 0.1$, $m_{BH} = 10$, $B_4 \approx 10^5$, and $0.3594 <
a_ * < 0.9980$, we have

\begin{equation}
\label{eq10} 7.3\times 10^{ - 16} < \mid n_{\delta}\mid  / \bar
{n}_p < 1.28\times 10^{ - 14}.
\end{equation}

\noindent It seems reasonable that the fluctuation in the number
densities of the accreting plasma can be realized in the realistic
astrophysical context for the small value of ${n_\delta }
\mathord{\left/ {\vphantom {{n_\delta } {\bar {n}_p }}} \right.
\kern-\nulldelimiterspace} {\bar {n}_p }$ given in equation
(\ref{eq10}).

Another issue related to the state transition in BHXBs is how to
estimate the timescale of the fluctuation in density of the
accreting plasma. If the toroidal current arises from the
fluctuation of the number density of the accreting plasma, we think
that the variation of the power-law index $n$ might occur due to
this fluctuation, and it gives rise to the state transition in
BHXBs. Not long ago, Brown et al. (2000) discussed the MC effect on
the accretion flow, and they estimated the viscous inflow time for
the fluctuations as

\begin{equation}
\label{eq11} \tau \sim r \mathord{\left/ {\vphantom {r {\upsilon _r
}}} \right. \kern-\nulldelimiterspace} {\upsilon _r } \sim \left( {r
\mathord{\left/ {\vphantom {r H}} \right. \kern-\nulldelimiterspace}
H} \right)^2\alpha _{vis}^{ - 1} \Omega _D^{ - 1} ,
\end{equation}

\noindent where $\upsilon _r $ is radial velocity of the accreting
plasma and $H$ is the height of the disk at radius $r$. We take the
coefficient of kinematic viscosity $\alpha _{vis} = 0.1$ in
calculations. Since the variation occurs within the outer boundary
of the MC, we calculate the timescale corresponding to $r_{out} =
\xi _{lower} r_{ms} $ by using equation (\ref{eq11}) as listed in
Table 3.

It is obvious, from equation (\ref{eq11}) and Table 3, that the
ratio $r \mathord{\left/ {\vphantom {r H}} \right.
\kern-\nulldelimiterspace} H$ is dominative in determining the
timescales of state transitions, which are insensitive to the
parameters, $m_{BH} $, $a_ * $ and $\xi _{lower} $. Inspecting Table
3, we find that the timescales of state transitions in BHXBs from
less than one second to more than one hour can be fitted by the
fluctuation in density of the accreting plasma. We expect that the
timescales of the state transitions in the above sources can be
fitted by adjusting the ratio $r \mathord{\left/ {\vphantom {r H}}
\right. \kern-\nulldelimiterspace} H$ based on this simplified
model.

Since the fluctuation in densities of the accreting plasma is
stochastic, this results in a stochastic variation of the power-law
index $n$, and it is consistent with the observation of the state
transitions in BHXBs: a hard state can transit to a soft one and
then back to the hard one again, passing across the jet line several
times as shown in Figure 7 of FBG04.

\section{DISCUSSION}

In this paper, we assume that Keplerian frequencies at two locations
are in the 3:2 ratio based on the CEBZMC model. Compared with the
Kluzniak-Abramowicz QPO resonance model, the 3:2 QPO pairs arising
from the two hotspots are produced by the MC between a rotating BH
and its surrounding disk. The BH spins of several BH sources
measured by different methods are compared. It turns out that the BH
spin of GRO J1655-40 measured by the CEBZMC model is in a good
agreement with the recent results based on X-ray continuum fitting.
In addition, the correlation of the 3:2 QPO pairs with the jet from
the BH systems including Sgr A* is discussed in the   parameter
space. It is shown that the `jet line' in HID can be interpreted
naturally by the CEBZMC model. Finally, we suggest that the state
transition in BHXBs could be realized by virtue of the variation of
the power-law index n, which could be related to the fluctuation of
the number densities of the accreting plasma from the companion
star.

In our model the 3:2 QPO pairs are determined by the BH spin $a_ * $
and the power-law index $n$ for the given BH mass. The parameter $n$
is introduced to describe the basic feature of the large-scale
magnetic field anchored at the disk, indicating the degree of its
concentration at the inner region. It turns out that the parameter
$n $ plays a very important role not only in fitting the 3:2 QPO
pairs but also in interpreting the state transitions of BHXBs.

It is easy to find from the parameter spaces in Figures 2---6 that
the state transition from a hard state to a soft state can be
realized by decreasing the BH spin, and the inverse transition
occurs with the increasing spin. However, our calculations show that
the timescale for the variation of the spin is too long to fit the
observations. In addition, the evolution of the BH spin is generally
one-direction, i.e., it decreases from a high spin to the
equilibrium spin or increases from a low spin to the equilibrium
spin as argued in W02. Thus, this account of the BH spin is not
consistent with the observations: the states of BHXBs can switch
from time to time between a hard state and a soft state. Compared
with the BH spin the variation of the parameter $ n$ involves the
timescale of the fluctuation of the number density, which is
consistent with the state transition of BHXBs both in the timescale
and in the repeating switches between the hard and soft states.

Very recently, Ma et al. (2006) introduced corona into the model of
CEBZMC, which might be helpful to understand the association of the
SPL state with the high frequency QPOs in BHXBs as argued in MR06.
We shall discuss this issue in our future work.

\acknowledgments
 {\bf Acknowledgements:}This work is supported by the National
Natural Science Foundation of China under grants 10373006, 10573006
and 10121503. We are very grateful to the anonymous referee for his
(her) instructive comments on the role of MC in the 3:2 frequency
ratio and the related issues.

\begin{figure}

\epsscale{0.5}
\begin{center}
\plotone{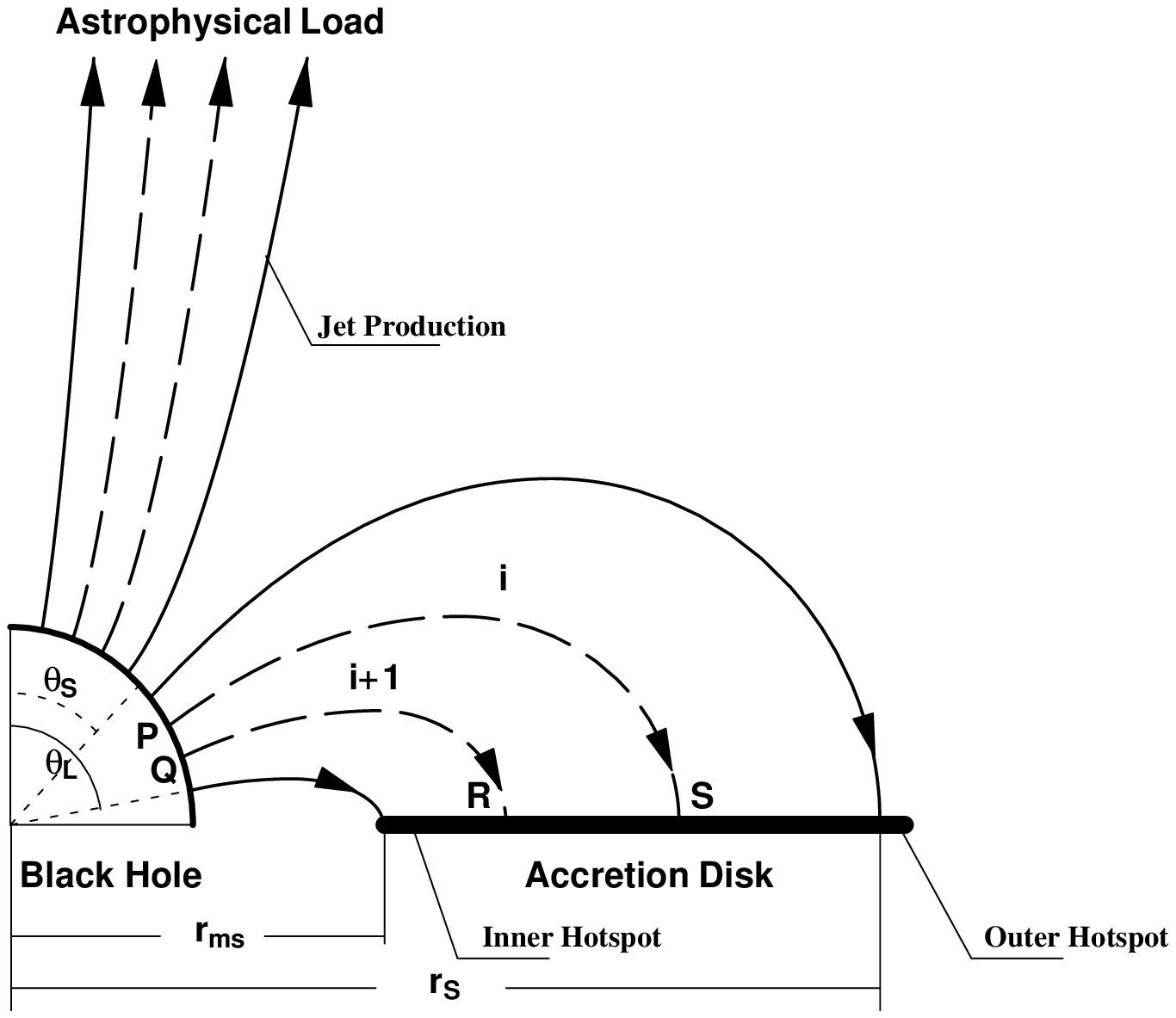}
\end{center}

\caption{Poloidal magnetic field connecting a rotating BH with a
remote astrophysical load and the surrounding disk. The inner and
outer hotspots are located at different places of the disk.}
\label{fig1}

\end{figure}

\begin{figure}
\epsscale{0.5}
\begin{center}
\plotone{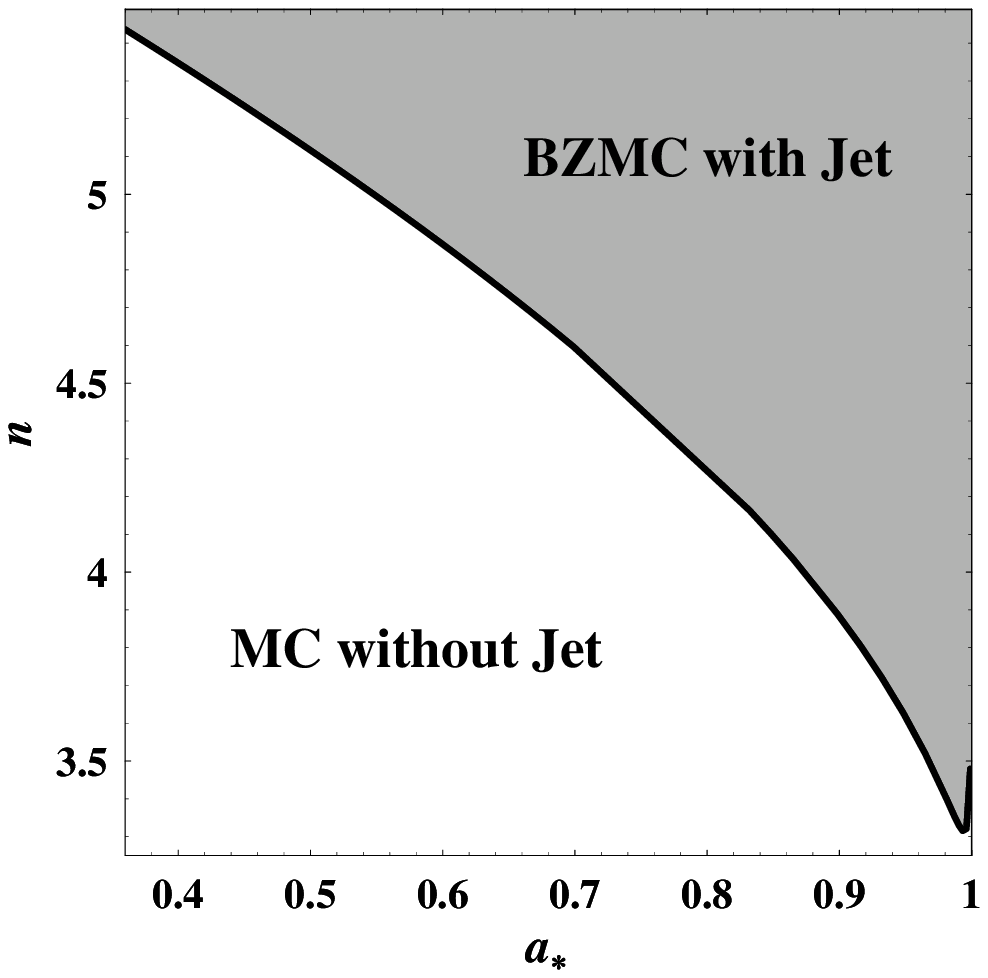}
\end{center}

\caption{The contour of angular boundary $\theta _S ( a_ * ,n ) = 0$
in $a_ * - n$ parameter space.} \label{fig2}
\end{figure}

\begin{center}
\begin{figure}
\epsscale{0.30}
\begin{center}
 \plotone{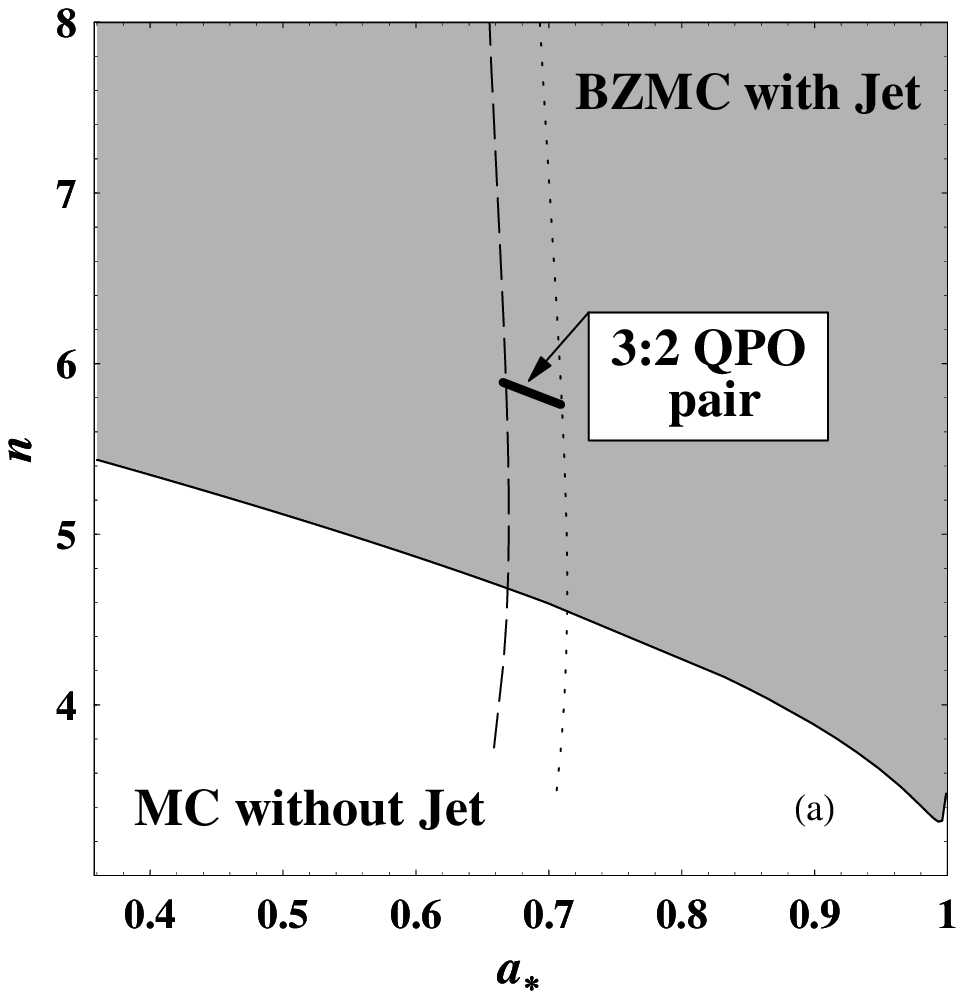}
\end{center}
\begin{center}
 \epsscale{0.30}
  \plotone{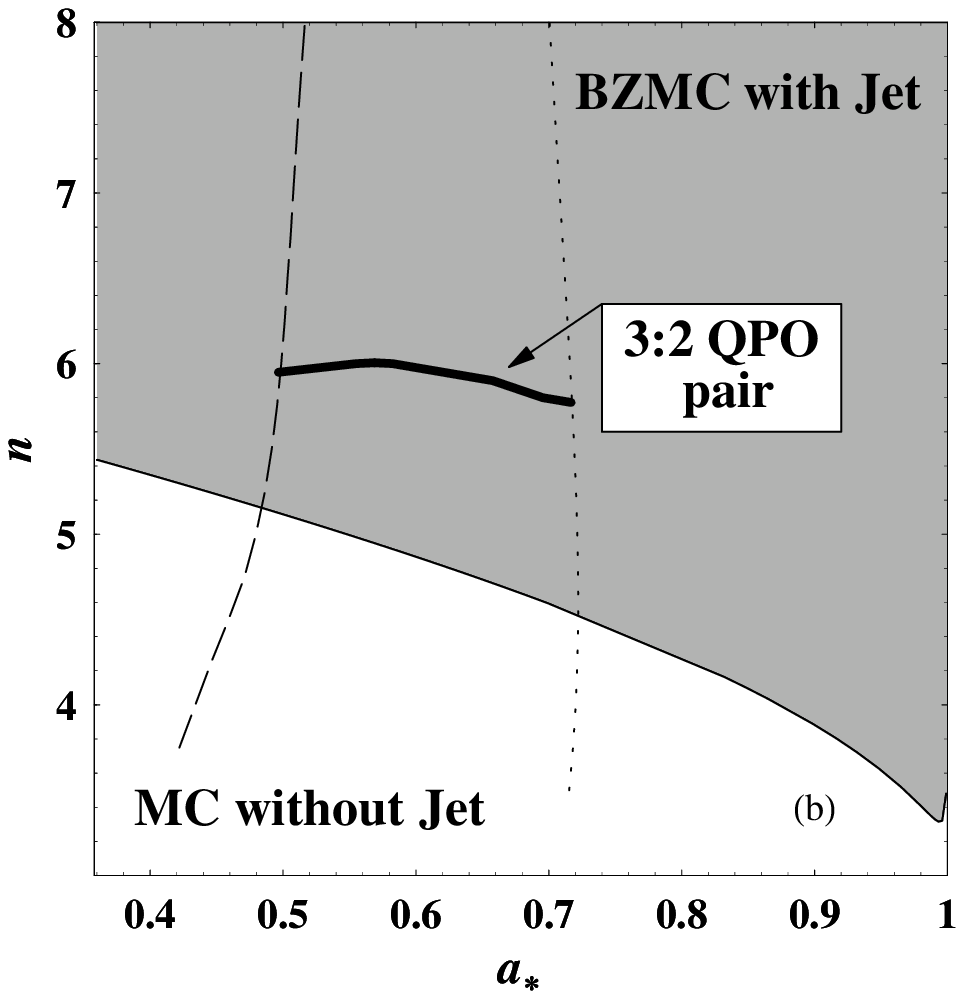}
\end{center}
\begin{center}
 \epsscale{0.30}
 \plotone{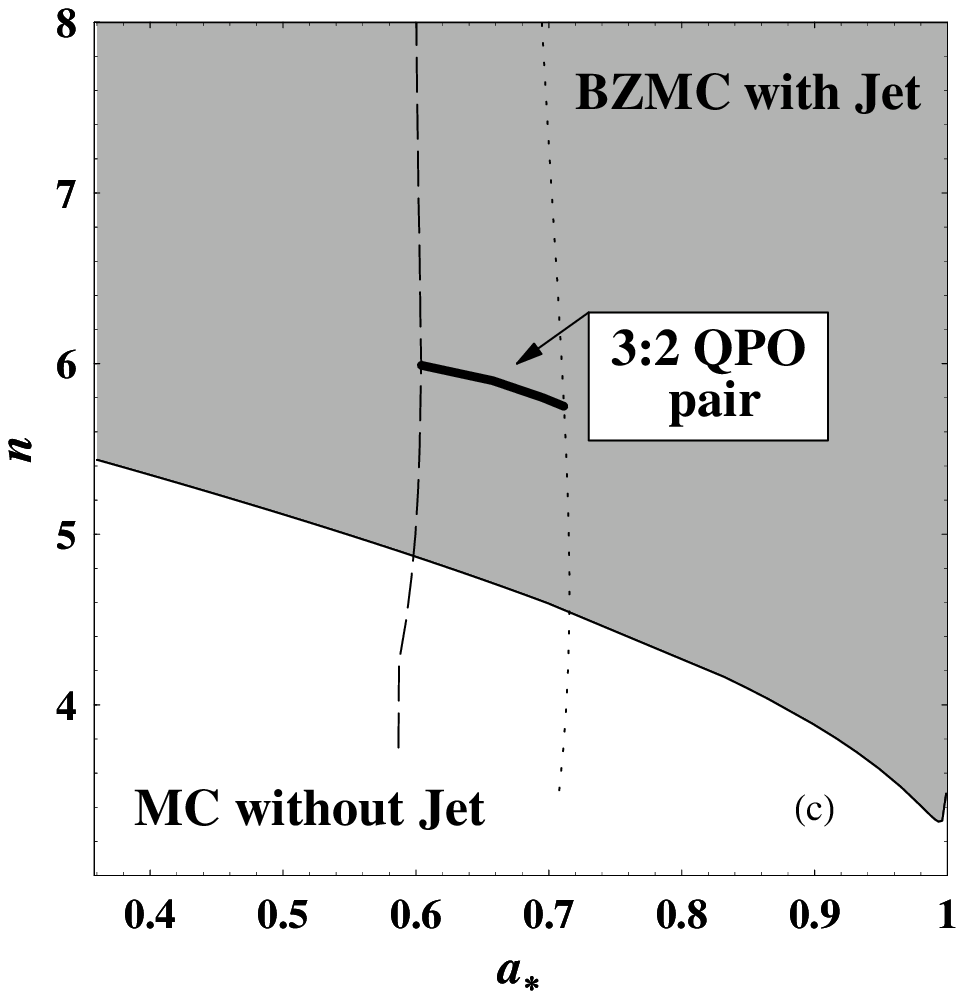}
 \end{center}
\begin{center}
 \epsscale{0.30}
 \plotone{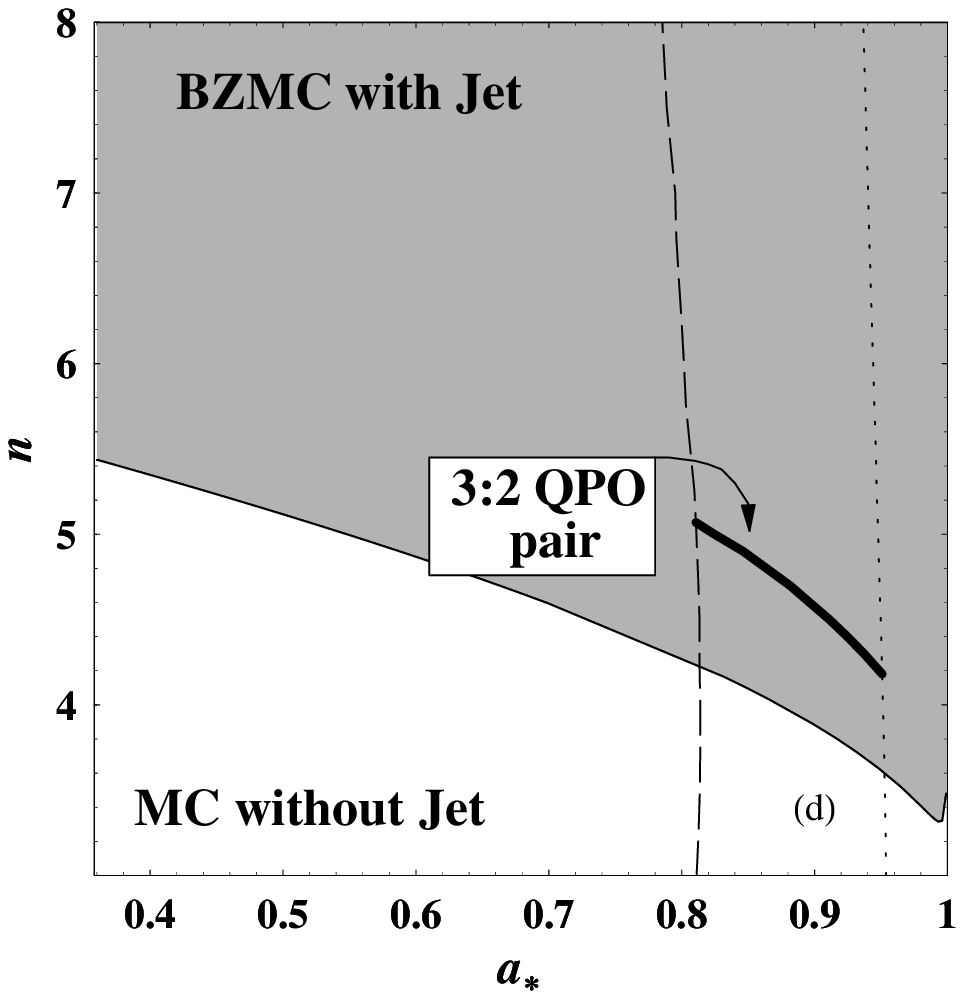}
 \end{center}
\caption{Two contours of $\nu _{upper} = const$ corresponding to the
lower and upper BH masses are shown respectively in dashed and
dotted lines, between which the thick solid line represents the BH
state with the 3:2 QPO pair for (a) GRO J1655-40, (b) XTEJ1550-564,
(c) GRS 1915+105 and (d) Sgr A*.} \label{fig3}
\end{figure}
\end{center}

\begin{figure}

\epsscale{0.5}
\begin{center}
\plotone{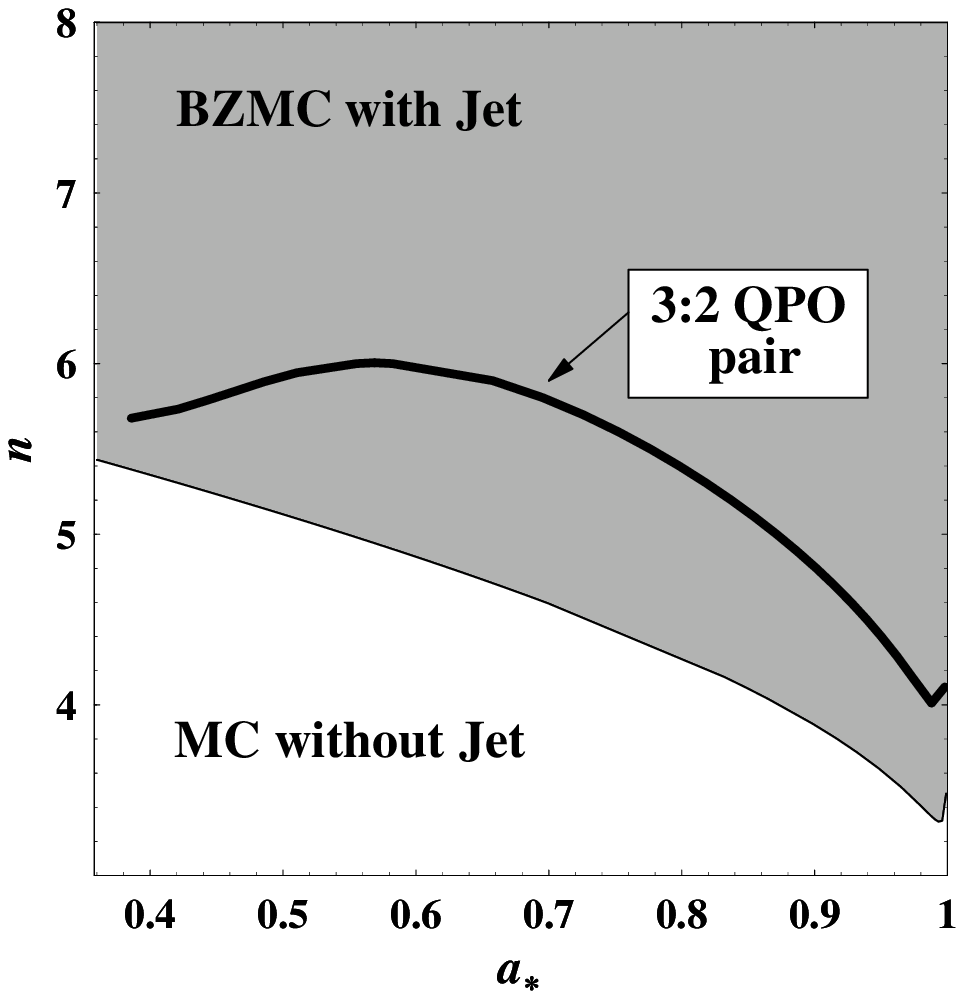}
\end{center}

\caption{The characteristic line of the 3:2 QPO pair (240, 160Hz)
for H1743-322 in $a_ * - n$ parameter space.} \label{fig4}

\end{figure}

\begin{figure}

\epsscale{0.5}
\begin{center}
\plotone{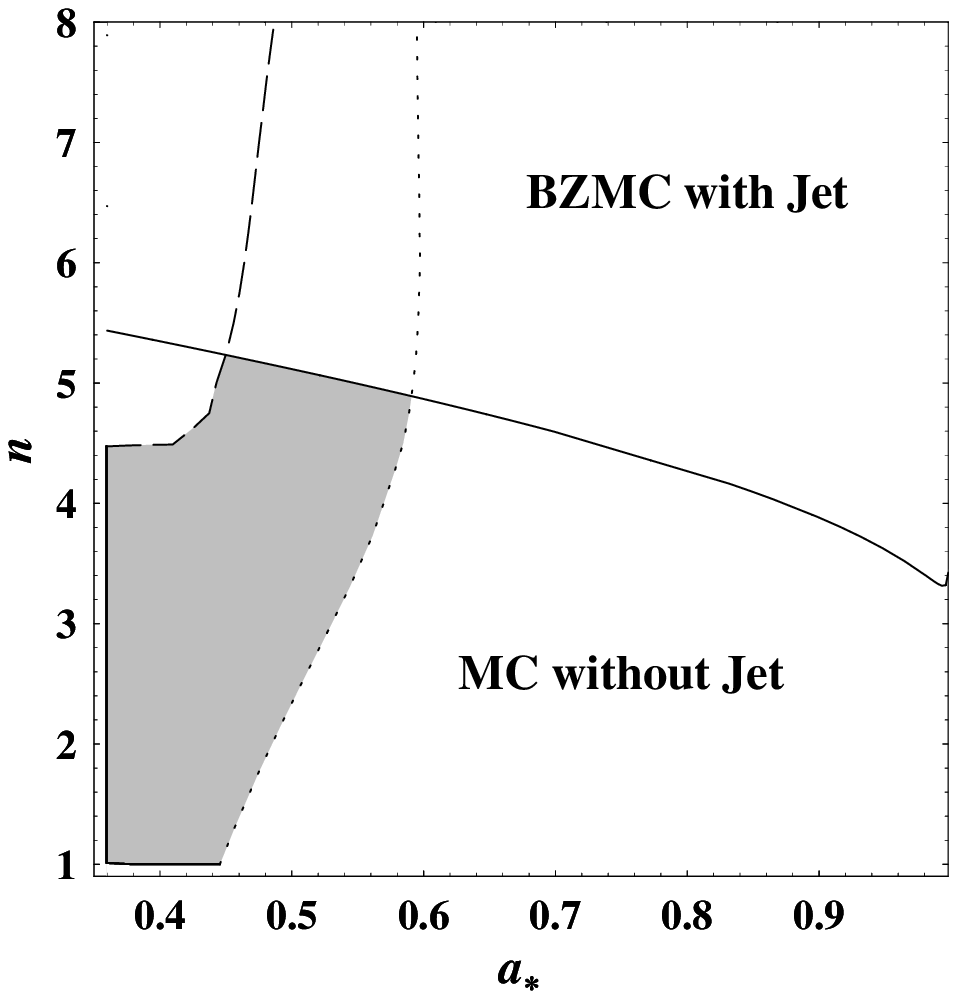}
\end{center}

\caption{Two contours of $\nu _{QPO} = 190Hz$ corresponding to the
lower and upper BH masses of XTE J1859+226 are shown in dashed and
dotted lines, respectively.} \label{fig5}

\end{figure}

\begin{figure}

\epsscale{0.5}
\begin{center}
\plotone{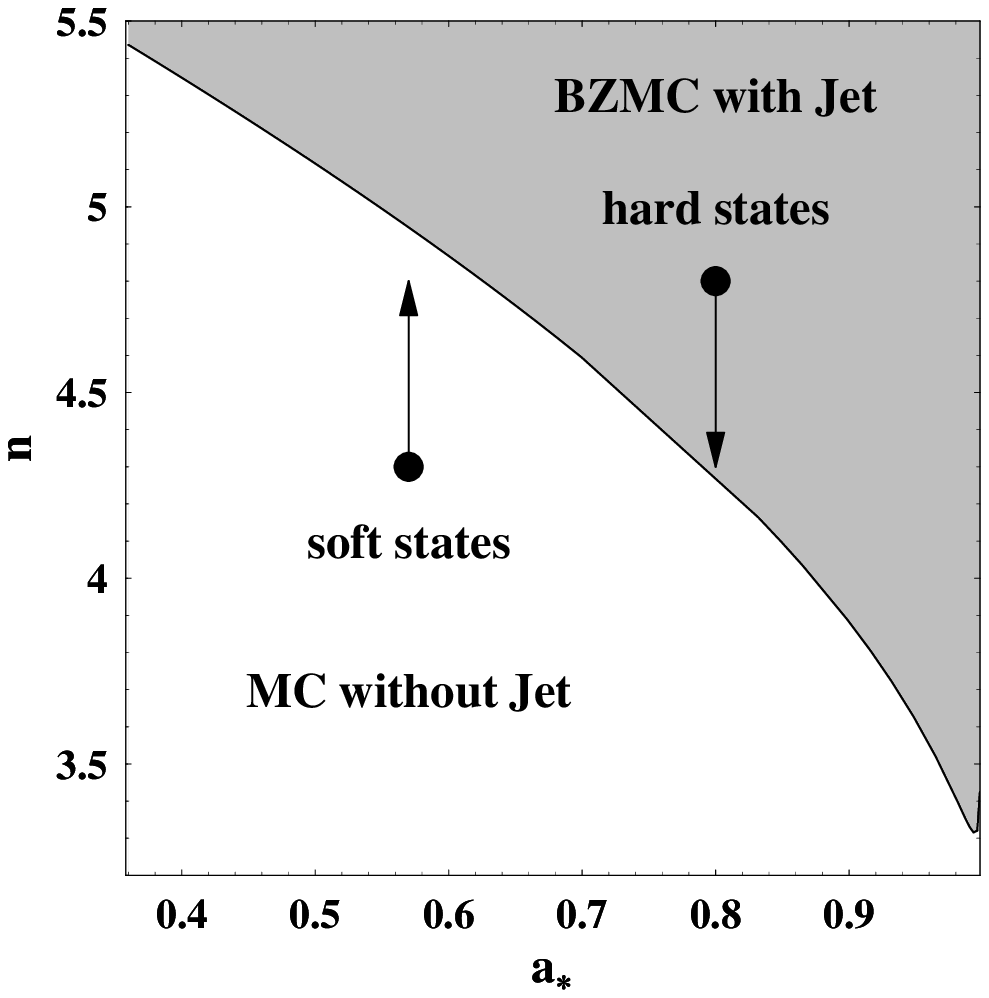}
\end{center}

\caption{A schematic drawing for interpreting the state transitions
of BHXBs in $a_ * - n$ parameter space.} \label{fig6}

\end{figure}


\clearpage

\begin{deluxetable}{ccccccccc}
\tabletypesize{\scriptsize} \tablecaption{The fitting results for
the 3:2 QPOs pairs in GRO J1655-40, XTE J1550-564, GRS 1915+105 and
Sgr A*.} \tablewidth{0pt} \tablehead{\colhead{Source} &
\colhead{$m_{_{BH}} $}& \colhead{$a_{*} $} & \colhead{$n$} &
\colhead{Jet} & \colhead{Inner Hotspot} & &\colhead{Outer Hotspot}&
} \startdata
 &
 &
 &
 &
 & $\xi _{upper} $ & $\nu _{upper} ( Hz )$ & $\xi _{lower} $ &
$\nu _{lower} ( Hz )$  \\
\hline
 GRO J1655-40& 6.0& 0.667&
 5.887&
 Yes&
1.421& 450& 1.887  &
300 \\
\cline{2-4} \cline{6-6} \cline{8-8}
& 6.6& 0.709&
 5.763&
 &
1.400& & 1.863  &
 \\
\hline
 XTE J1550-564& 8.4& 0.603&
 5.996&
 Yes&
1.473& 276& 1.949  &
184 \\

\cline{2-4} \cline{6-6} \cline{8-8}
& 10.8& 0.710&
 5.771&
 &
1.397& & 1.860  &
 \\
 \hline
 GRS 1915+105& 10.0& 0.496&
 5.955&
 Yes&
1.660& 168& 2.174  &
113 \\
\cline{2-4} \cline{6-6} \cline{8-8}

& 18.0& 0.716&
 5.785&
 &
1.395& & 1.845  &
 \\
 \hline
 Sgr A*& 2.6x10$^{6}$& 0.811&
 5.069&
 Yes&
1.378& 1.445 x10$^{ - 3}$& 1.961  &
0.886 x10$^{ - 3}$ \\

\cline{2-4} \cline{6-6} \cline{8-8}

& 4.4x10$^{6}$& 0.951&
 4.181&
 &
1.334& & 1.957  &
 \\
\hline
\enddata

\tablecomments{The value ranges of the BH mass corresponding to GRO
J1655-40, GRS 1915+105 and XTE J1550-564 are adopted from RM06, and
the BH mass of Sgr A* is taken from T$\ddot{o}$r$\ddot{o}$k (2005).}
\end{deluxetable}

\begin{deluxetable}{cccc}
\tabletypesize{\scriptsize} \tablecaption{The BH spins measured by
X-ray continuum and 3:2 QPO pairs} \tablewidth{0pt}
\tablehead{\colhead{Sources} & & \colhead{Methods} &  } \startdata

 & X-ray continuum& 3:2 QPO pairs&
 \\
\cline{3-4}
&
 &
ERM& CEBZMC
\\
\hline GRO J1655-40& 0.633-0.651 (D06), 0.65-0.80 (M06) & 0.2-0.67
(AK01), 0.96 (AK04), & 0.667-0.709
\\
&  0.65-0.75 (S06), 0.7-0.95 (Z97),  & 0.996 (A04b), 0.31-0.42
(KF06),  &
\\
&0.68-0.88 (G01)&  0.64-0.76 (Br05)&
\\
\hline

XTE J1550-564& 0.71-0.87 (D06)& 0.94 (AK04), 0.99616 (A04b) &
0.603-0.710
\\
& &  0.11-0.42 (KF06), 0.1-0.6 (RM02), &
\\
& &  0.41-0.77 (Br05)&
\\
\hline

 GRS 1915+105&
$>$0.98 (M06), 0.7 (MD06)& 0.84(AK04), 0.996 (A04b), & 0.496-0.716
\\
& & negative-0.44(KF06), -0.09-0.78(Br05)&
\\
\hline

Sgr A*& & 0.99616 (A06), 0.9865-0.9965 (A04a)&
0.811-0.951 \\
\hline

\enddata

\tablecomments{\textbf{The abbreviations for the references in Table
2 are given as follows:} \\AK01---Abramowicz {\&} Kluzniak (2001);
AK04---Abramowicz {\&} Kluzniak (2004); A04a---Aschenbach, et al.,
(2004); A04b---Aschenbach, (2004); A06---Aschenbach, (2006);
Br05---Bursa, (2005); D06---Davis et al. (2006); G01---Gierlinski et
al.(2001); KF06---Kato {\&} Fukue (2006); M06
---McClintock et al.(2006); MD06 --- Middleton et al. (2006); RM02---
Remillard et al. (2002); S06---Shafee et al.(2006); Z97---Zhang et
al. (1997).}
\end{deluxetable}

\begin{deluxetable}{ccccccc}
\tabletypesize{\scriptsize} \tablecaption{The timescale of state
transition in BHXBs fitted by the fluctuations in accreting plasma.}
\tablewidth{0pt} \tablehead{\colhead{Sources} & \colhead{$m_{_{BH}}
$}& \colhead{$a_ * $} & \colhead{$\xi _{lower} $} & \colhead{} &
\colhead{$\tau $(sec)} &\colhead{}}

\startdata
 &
 &
 &
 & $r/H = 10$ & $r/H = 100$ & $r/H = 1000$
 \\
\hline

GRO J1655-40& 6.0& 0.667& 1.887& 0.5305& 53.05&
5305 \\
\cline{2-4}
 &
6.6& 0.709& 1.863&
 &
 &
  \\
\hline XTE J1550-564& 8.4& 0.603& 1.949& 0.8650& 86.50&
8650 \\
\cline{2-4}
 &
10.8& 0.710& 1.860&
 &
 &
  \\
\hline GRS 1915+105& 10.0& 0.496& 2.174& 1.408& 140.8&
14080 \\
\cline{2-4}
 &
18.0& 0.716& 1.845&
 &
 &
  \\
\hline Sgr A*& 2.6x10$^{6}$& 0.811& 1.961& $1.796\times 10^5$&
$1.796\times 10^7$&
$1.796\times 10^9$ \\
\cline{2-4}
 &
4.4x10$^{6}$& 0.951& 1.957&
 &
 &
  \\
\hline

\enddata

\end{deluxetable}


\begin{thebibliography}

\bibitem[1]{b1}{Abramowicz, M. A., {\&} Kluzniak, W., 2001, A{\&}A, 374, L19
(AK01)}

\bibitem[1]{b1}{Abramowicz M.A., Bulik T., Bursa M. \& Kluzniak W., 2003, A{\&}A
Letters 404, L21}

\bibitem[1]{b1}{Abramowicz, M. A., {\&} Kluzniak, W., 2004, in AIP Conf.
Proceedings, 714, \textit{X-ray Timing 2003: Rossi and Beyond,} ed.
P Kaaret, F K. Lamb, J H. Swank. (NY: AIP), 21 (AK04)}


\bibitem[1]{b1}{Abramowicz M.A., 2005, (ed.), 2005, AN, Vol. 326, No. 9 (Abramowicz 2005)}

\bibitem[1]{b1}{Abramowicz M.A., Barret D., Bursa M., Horak J. Kluzniak W.
Olive J.-F. Rebusco, P. {\&} T$\ddot{o}$r$\ddot{o}$k G., 2005, AN,
326, 864}



\bibitem[1]{b1}{Aschenbach, B., et al. 2004a, A{\&}A, 417, 71
(A04a)}

\bibitem[1]{b1}{---. 2004b, A{\&}A, 425, 1075 (A04b)}

\bibitem[1]{b1}{---. 2006, Chinese J. Astron. Astrophys. Suppl., 6, 221 (A06)}

\bibitem[1]{b1}{Belloni, T., Mendez, M., King, A. R., van der Klis, M., {\&} van
Paradijs, J. 1997a, ApJ, 479, L145}

\bibitem[1]{b1}{---.1997b, ApJ, 488, L109}

\bibitem[1]{b1}{Belloni, T., Klein-Wolt, M., Mendez, M., van der Klis, M., {\&} van
Paradijs, J. 2000, A{\&}A, 355, 271}

\bibitem[1]{b1}{Belloni T., Mendez M. {\&} Homan J., 2005, A{\&}A, 437,
209}

\bibitem[1]{b1}{Belloni, T., 2006, Adv. Space Res., 38, 2801 (B06)}

\bibitem[1]{b1}{Blandford, R. D., {\&} Znajek, R. L. 1977, MNRAS, 179,
433}

\bibitem[1]{b1}{Blandford, R. D., 1999, in ASP Conf. Ser. 160, \textit{Astrophysical
Discs}: An EC Summer School,

     ed. J. A. Sellwood {\&} J. Goodman (San Francisco: ASP), 265}

\bibitem[1]{b1}{Blandford, R. D., 2002, Lighthouses of the Universe: The Most
Luminous Celestial Objects and Their Use for Cosmology Proceedings
of the MPA/ESO/, p. 381.}

\bibitem[1]{b1}{Bradt, H. V., Rothschild, R. E., {\&} Swank, J. H., 1993, A{\&}AS,
97, 355}

\bibitem[1]{b1}{Brown, G. E., et al. 2000, New Astronomy 5, 191}

\bibitem[1]{b1}{Bower, G. C., Falcke, H., Herrnstein, R. M., et al. 2004, Science,
304, 704}

\bibitem[1]{b1}{Bursa, M., 2005, in Proceedings of RAGtime 6/7: Workshops on black
holes and neutron stars,

    ed. S. Hled\'{\i}k {\&} Z. Stuchl\'{\i}k (Silesian University in
Opava, Czech), 39 (Br05)}

\bibitem[1]{b1}{Davis, S. W., Done, C., {\&} Blaes, O. M., 2006, ApJ, 647, 525
(D06)}

\bibitem[1]{b1}{Fender, R.P., Belloni, T., {\&} Gallo, E., 2004, MNRAS, 355, 1105
(FBG04)}

\bibitem[1]{b1}{Gierlinski, et al. 2001, MNRAS, 325, 1253 (G01)}

\bibitem[1]{b1}{Homan, J., et al. 2005, ApJ, 623, 383}

\bibitem[1]{b1}{Kalemci, E. et al. 2006, ApJ, 639, 340}

\bibitem[1]{b1}{Kato, S., {\&} Fukue, J., 2006, PASJ, 58, 909 (KF06)}

\bibitem[1]{b1}{Lachowicz P., Czerny B. {\&} Abramowicz M.A., 2006,
astro-ph/0607594}


\bibitem[1]{b1}{Lamb F. K. {\&} Miller M.C., 2004, Bull. AAS, 36, 937}



\bibitem[1]{b1}{Li, L. -X., 2000, ApJ, 533, L115}

\bibitem[1]{b1}{---. 2002, ApJ, 567, 463}

\bibitem[1]{b1}{Livio, M., Pringle, J. E., {\&} King, A. R., 2003, ApJ, 593, 184
(LPK03)}

\bibitem[1]{b1}{Ma, R.-Y., Wang, D.-X., {\&} Zuo, X.-Q., 2006, A{\&}A, 453,
1}

\bibitem[1]{b1}{McClintock, J E, {\&} Remillard R A 2006. In Compact Stellar X-ray
Sources, ed. WHG Lewin, M van der Klis, pp. 157--214. Cambridge:
Cambridge University Press. (astro-ph/0306213) (MR06)}

\bibitem[1]{b1}{McClintock, J E et al., 2006, ApJ, 652, 518 (M06)}

\bibitem[1]{b1}{Middleton, M., et al., MNRAS, 373, 1004 (MD06)}

\bibitem[1]{b1}{Mirabel, I. F., {\&} Rodriguez L. F., 1998, Nat, 392,
673}

\bibitem[1]{b1}{---. 1999, ARA{\&}A, 37, 409}

\bibitem[1]{b1}{Miller, J. M. et al. 2001, ApJ, 563, 928}

\bibitem[1]{b1}{Remillard, R. A., et al., 2002, ApJ, 564, 962}

\bibitem[1]{b1}{Remillard, R. A., {\&} Muno, M. P., ApJ, 2002, 580, 1030
(RM02)}

\bibitem[1]{b1}{Remillard R.A., 2005, AN, 326, 804}

\bibitem[1]{b1}{Remillard, R. A., et al. 2006, ApJ, 637, 1002}

\bibitem[1]{b1}{Remillard, R. A., {\&} McClintock J. E., 2006, ARA{\&}A, 44, 49
(RM06)}

\bibitem[1]{b1}{Shafee, R., et al. ApJ, 2006, 636, L113 (S06)}

\bibitem[1]{b1}{Stella L. {\&} Vietri M., 1999, Phys. Rev. Lett., 82,
17}

\bibitem[1]{b1}{Strohmayer, T. E., 2001a, ApJ, 552, L49}

\bibitem[1]{b1}{---. 2001b, ApJ, 554, L169}

\bibitem[1]{b1}{Thorne, K. S. 1974, ApJ, 191, 507}

\bibitem[1]{b1}{T$\ddot{o}$r$\ddot{o}$k G., Abramowicz M. A., Kluzniak, W., Stuchl¨ªk, Z., 2005,
A{\&}A, 436, 1}

\bibitem[1]{b1}{T$\ddot{o}$r$\ddot{o}$k, G., 2005a, AN, 326, 856}

\bibitem[1]{b1}{T$\ddot{o}$r$\ddot{o}$k G., 2005b, A{\&}A, 440, 1}

\bibitem[1]{b1}{Wagoner, R. V., Silbergleit, A. S., {\&} Ortega-Rodriguez, M. 2001,
ApJ, 559, L25}

\bibitem[1]{b1}{Wang, D.-X., Xiao K., {\&} Lei W.-H. 2002, MNRAS, 335, 655
(W02)}

\bibitem[1]{b1}{Wang, D.-X., et al., 2003a, ApJ, 595, 109 (W03)}

\bibitem[1]{b1}{---. 2003b, MNRAS, 344, 473}

\bibitem[1]{b1}{---. 2004, ApJ, 601, 1031 (W04)}

\bibitem[1]{b1}{Wang, et al., MNRAS, 2005, 359, 36 (W05)}

\bibitem[1]{b1}{Zhang, S. N., Cui, W., {\&} Chen, W., 1997, ApJ, 482, L155
(Z97)}

\end{thebibliography}
\end{document}